\documentclass[12pt,a4paper]{article}
\usepackage{graphicx}
\usepackage{wrapfig}
\usepackage{slashed}

\newcommand\pubnumber{}
\newcommand\pubdate{}

\def\Title#1{\begin{center} {\Large #1 } \end{center}}
\def\Author#1{\begin{center}{ \sc #1} \end{center}}
\def\Address#1{\begin{center}{ \it #1} \end{center}}

\newcommand\pubblock{\rightline{\begin{tabular}{l} \pubnumber\\
         \pubdate  \end{tabular}}}
\newenvironment{Abstract}{\begin{quotation}  }{\end{quotation}}
\newenvironment{Presented}{\begin{quotation} \begin{center}
             PRESENTED AT\end{center}\bigskip
      \begin{center}\begin{large}}{\end{large}\end{center} \end{quotation}}

\def\beq{\begin{equation}}
\def\eeq#1{\label{#1}\end{equation}}
\def\eeqn{\end{equation}}

\def\beqa{\begin{eqnarray}}
\def\eeqa#1{\label{#1}\end{eqnarray}}
\def\eeqan{\end{eqnarray}}

\let\bar=\overbar

\def\Dslash{\not{\hbox{\kern-4pt $D$}}}
\def\dslash{\not{\hbox{\kern-2pt $\del$}}}

\def\msb{{\bar{\ssstyle M \kern -1pt S}}}

\begin{document}
\begin{titlepage}
\pubblock

\vfill
\Title{Kinematical Signatures of $W^+$ Pair Production}
\vfill
\Author{Miroslav Myska}
\Address{Czech Technical University in Prague, FNSPE, Brehova 7, 115 19, Prague, \\Czech Republic}
\vfill
\begin{Abstract}
The underlying event measurement may be crucial for the new physics
searches at high energies at LHC. This study presents the influence
of double parton scattering on the same-sign di-muon production,
where only the positive charge is taken into account for now. The
signal production cross section is found to be 0.94 $fb$ within the
kinematic range of the ATLAS detector. This creates around 25 per
cent of the total cross section for the searched final state.
\end{Abstract}
\vfill
\begin{Presented}
MPI@LHC 2010\\
2nd International Workshop on Multiple Partonic Interactions at the LHC\\
Glasgow, UK, November 29 -- December 3, 2010
\end{Presented}
\vfill
\end{titlepage}
\def\thefootnote{\fnsymbol{footnote}}
\setcounter{footnote}{0}

\section{Introduction}

Multiple Parton Interactions (MPI) are usually considered as a
perfect tool how to describe the multi-jet production at hadron
colliders, where a statistical processing plays the most important
role. Measurements at SPS \cite{PLB268_145} and at Tevatron
\cite{PRD56_3811} provided first accurate insights into the
phenomenon and studied its main signatures and cross section
magnitudes. Contemporary minimum bias measurements at LHC aim for
further tuning of Monte Carlo MPI models and all the subject becomes
still more and more significant.

The energy of 14 TeV designed for LHC could also allow the
measurements of much rare processes like vector boson pair
production. In this report, pair of positively charged W bosons
produced via MPI mechanism and decaying into pair of same-sign muons
is studied in detail. Herwig++ program \cite{EPJC58_639} is used to
prepare these proton-proton signal events while the physics
background processes are generated using the MadGraph's
\cite{JHEP09_028} matrix elements within the Herwig++ hadronization
and shower tools. Single Parton Scattering (SPS) production of
$W^+W^+jj$, $W^+Z$, $ZZ$, and $t\bar{t}$ is studied in order to find
the kinematical selection criteria for the best signal separation.

\section{$W^+W^+$ Signal Generation in Herwig++}

\begin{wrapfigure}{l}{6cm}
\begin{center}
\vspace{-0.5cm}
\includegraphics[width=4.5cm]{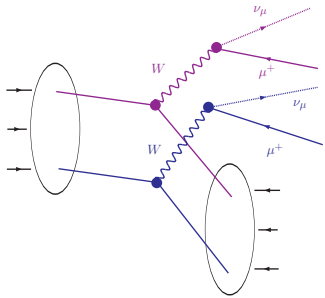}
\caption{Schematic diagram of $W^+$ pair production via
two independent parton annihilations.}\label{fig:WW}
\end{center}
\end{wrapfigure}

Physics motivation is to perform search for
MPI effects in the new channel that has
never been measured yet. The vector boson pair production is chosen
as a very straightforward test of the current stage of the MPI models.
The di-muon MPI signal process (see Fig. \ref{fig:WW}) consists of two hard and independent $W^+$
creations that allow only six flavor combinations of annihilating quark-antiquark
pairs:
\begin{equation}
u\bar{d} (75.9\%),~u\bar{s} (19.4\%),~u\bar{b} (3.4\%),~c\bar{d} (1.3\%),~c\bar{s}
,~c\bar{b} \rightarrow W^+ \rightarrow \mu^+ \nu_{\mu}.
\end{equation}

\noindent The numbers in brackets indicate the individual contributions to the matrix element
for the proton-proton interaction at $\sqrt{s}$ = 14 TeV using the CTEQ6L1 PDFs \cite{JHEP07_012}.
The precise measurement of the cross section could bring new constraints on the geometrical
coefficients $\Theta$ \cite{PRD63_057901} characterizing the prediction that different partons are distributed
inside the hadron according to the different transverse distribution functions. The overlap
function thus needs to carry two additional indices for interacting parton flavors and
the cross section can be written as

\begin{equation}
\sigma_D = \frac{1}{2} \sum_{ijkl}\int \sigma_S^{ij}\sigma_S^{kl}
A^i_k(b)A^j_l(b) d^2b= \frac{1}{2} \sum_{ijkl}
\sigma_S^{ij}\sigma_S^{kl} \Theta^{ij}_{kl}.
\label{eq:geom_coef}
\end{equation}

The normalization of the signal process cross section is done using
the widely accepted parameter called effective cross section,
$\sigma_{eff}$. Its value of 11.5 $mb$ used in

\noindent this study is derived from the CDF measurement \cite{PRD56_3811} and by applying some corrections based on
the considerations in \cite{PRD76_076006}. The final formula for sinal cross section calculation is

\begin{equation}
\sigma_D = \frac{\sigma_S^2}{2\sigma_{eff}},
\end{equation}

\noindent where $\sigma_S$ is the usual cross section for the single
$pp\rightarrow W^+ \rightarrow \mu^+\nu_{\mu}$ production.

Double Parton Scattering (DPS) term in this study denotes all the
proton-proton collisions containing at least two mentioned $W^+$
creations. This inclusive definition allows the presence of many
other parton sub-processes within the same collision. Herwig++
imitates this situation very well by generating additional QCD 2
$\rightarrow$ 2 parton interactions following the two main hard
sub-processes. Number of these sub-processes is pre-sampled
according to the Poissonian distribution calculated at the beginning
of the generation. The inclusivity of the double parton process is
satisfied almost perfectly because the chance of having the third
$W^+$ creation in the same collision is very small.

\begin{figure}[h!]
\begin{center}
 \begin{tabular}{cc}
  \begin{minipage}{.5\hsize}
   \begin{center}
     \includegraphics[width=5cm]{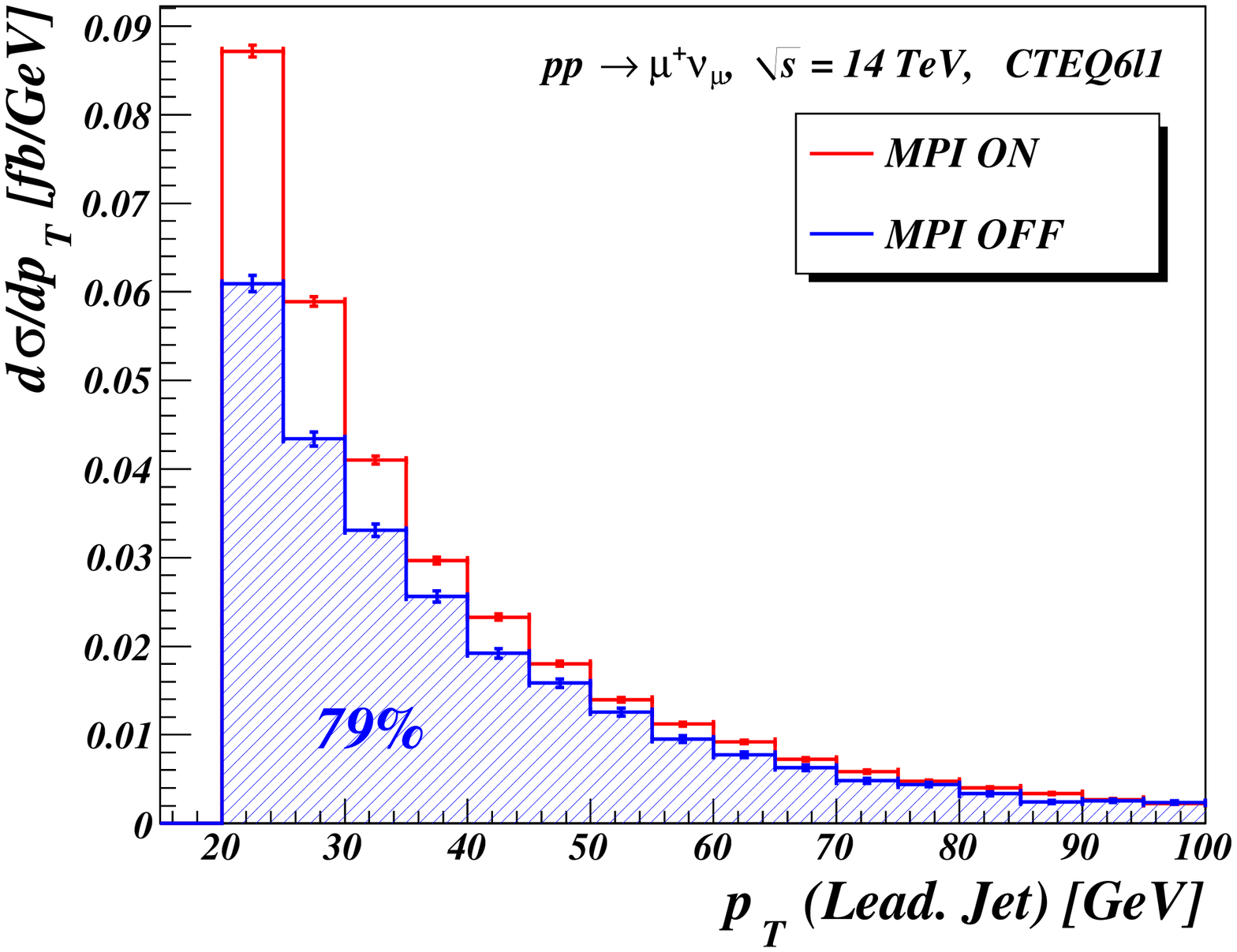}
   \end{center}
  \end{minipage}

  \begin{minipage}{.5\hsize}
   \begin{center}
     \includegraphics[width=5cm]{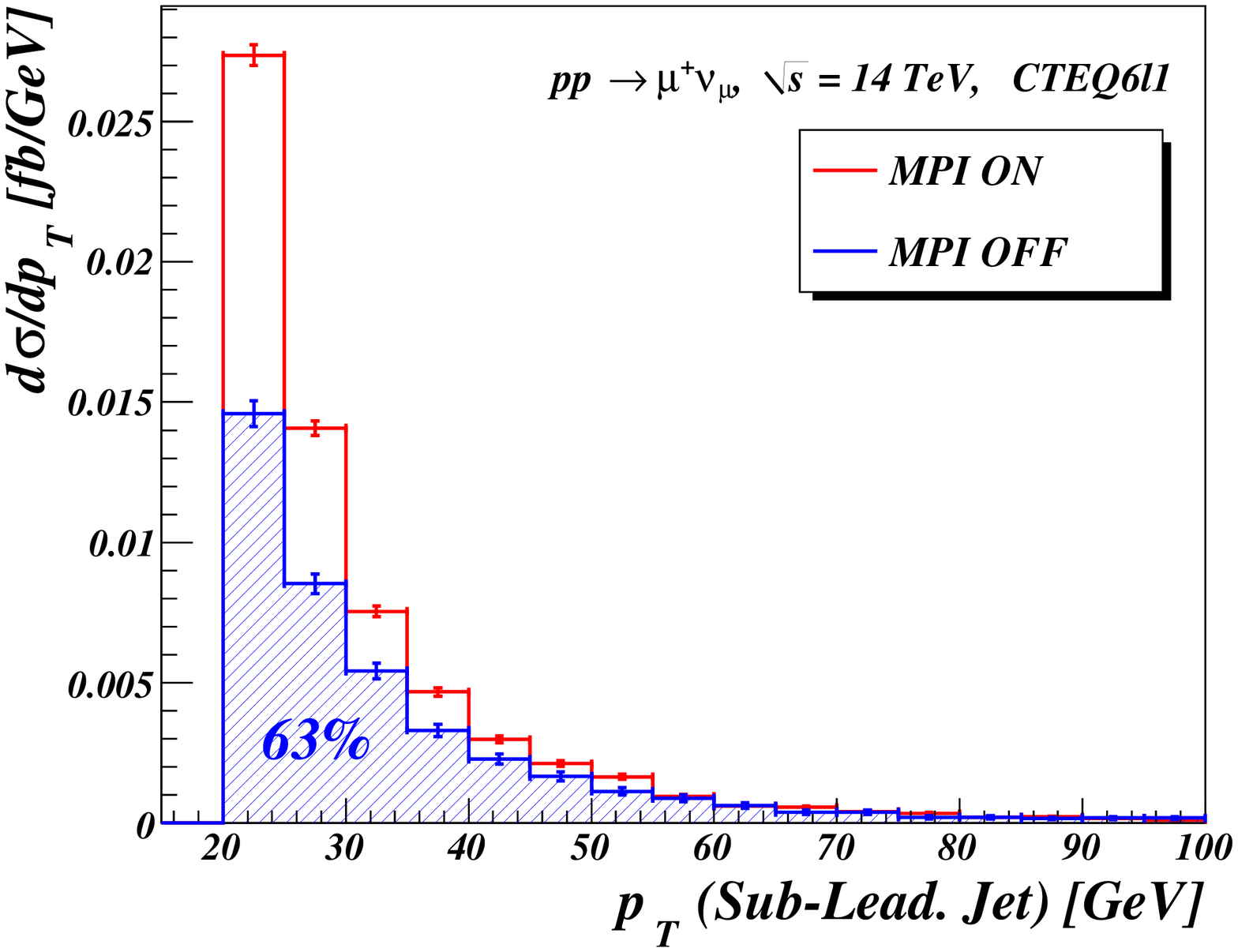}
   \end{center}
  \end{minipage}
 \end{tabular}
\caption{Transverse momentum distributions of the
leading (left plot) and the sub-leading jet (right plot) using
Herwig++ with MPI switched ON (red line) and OFF (blue line). The increase of
the MPI fraction with decreasing jet $p_T$ threshold is given by
relatively high number of additional QCD scatterings at semi-hard
level while the radiations from the primary process decrease rapidly
with its $p_T$.}
\label{fig:J_all_pt}
\end{center}
\end{figure}

These additional QCD parton sub-interactions are also of our interest. They contribute to the
underlying event and increase the amount of measured jets in the final state. For illustration,
plots in Fig. \ref{fig:J_all_pt} display the transverse
momentum distributions for the hardest (leading) and for the second
hardest (sub-leading) jet assuming that the event contains at least
one jet (two jets in the case of the sub-leading jet distribution)
with the $p_T^{jet} > 20~GeV$ and $|\eta^{jet}| < 4.5$. This
threshold was set according to the expected performance of the ATLAS
detector as an example of the sensible selection \cite{EPJC71_1}.
The $anti-k_t$ clustering algorithm \cite{JHEP04_063} implemented in
the FastJet package version 2.4.2 \cite{PLB641_57} was applied on
the non-lepton final state particles with the radius-like parameter
$R=0.4$.

\section{Selection Criteria and Background Supression}

Physics background processes that could result in the final state
created also by the pair of positively charged muons are studied.
Background steaming from the detector-related effects (like lepton
misidentification) still needs to be investigated further. Processes
taken into account are Single Parton Scattering (SPS) vector boson
pair production and heavy flavor quark-antiquark pair production.
The former case contains all three charge combinations $W^+W^+jj$,
$W^+Z$, and $ZZ$, where $Z$ denotes full matrix element for neutral
boson exchange. The latter source of the same-sign

\begin{wrapfigure}{l}{6cm}
\begin{center}
\vspace{-0.5cm}
\includegraphics[width=5cm]{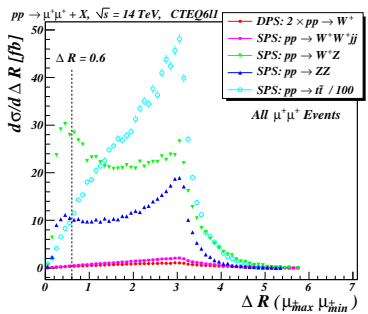}
\caption{Differential cross section as a function of the isolation distance
$\Delta R$ in the $\phi-\eta$ plane between two hardest positively charged muons.}
\label{fig:LepPair_dR}
\end{center}
\end{wrapfigure}

\noindent muon pairs is represented here by $t\bar{t}$ production,
where the huge cross section leaves a lot of space for the muon
radiation from short-living hadrons together with the heavy quark
decay to hard muon.

$t\bar{t}$ events were prepared using only Herwig++. The matrix
elements for the vector boson pair production were prepared using
MadGraph/MadEvent generator and were re-processed by Herwig++ in
order to complete the full proton-proton collision. Parton level
cuts were set as loose as possible only with one exception. Muons
coming from the processes with virtual $\gamma$ decays may propagate
very close to each other, see Fig. \ref{fig:LepPair_dR}. The $W^+Z$
and $ZZ$ productions thus had to be filtered at the parton level in
order to avoid the divergences in the cross section for too much
collinear muons. The edge close to the zero is fuzzy-distributed due
to the shower algorithm applied on the matrix element. The
reasonable choice of the minimal relative distance between the two
hardest muons in the $\eta-\phi$ plane seems to be the value of
$0.6$. Further, both muons had to satisfy the minimal requirements:
$p_T (\mu^+) > 5~GeV,|\eta (\mu^+)| < 2.5$ that are motivated by the
ATLAS detector performance \cite{CERN-OPEN-2008-020}. The
pseudorapidity cut is chosen according to the combined acceptance of
the inner tracking detector and the outer muon spectrometer.

DPS events analyzed in this study are characterized only by pair of
same-sign muons with tracks completely uncorrelated in any
projection. The two neutrinos from $W$ decays do not contribute to
the missing transverse energy in any significant way in comparison
to SPS $W^+Z$ production. The produced neutrino also rule out the
possibility of $W$ decay plane determination. The relative position
of the two planes from separate parton sub-interactions is usually
the strongest signature of the multiple parton scattering.

\begin{figure}[h!]
\begin{center}
 \begin{tabular}{cc}
  \begin{minipage}{.45\hsize}
   \begin{center}
     \includegraphics[width=6cm]{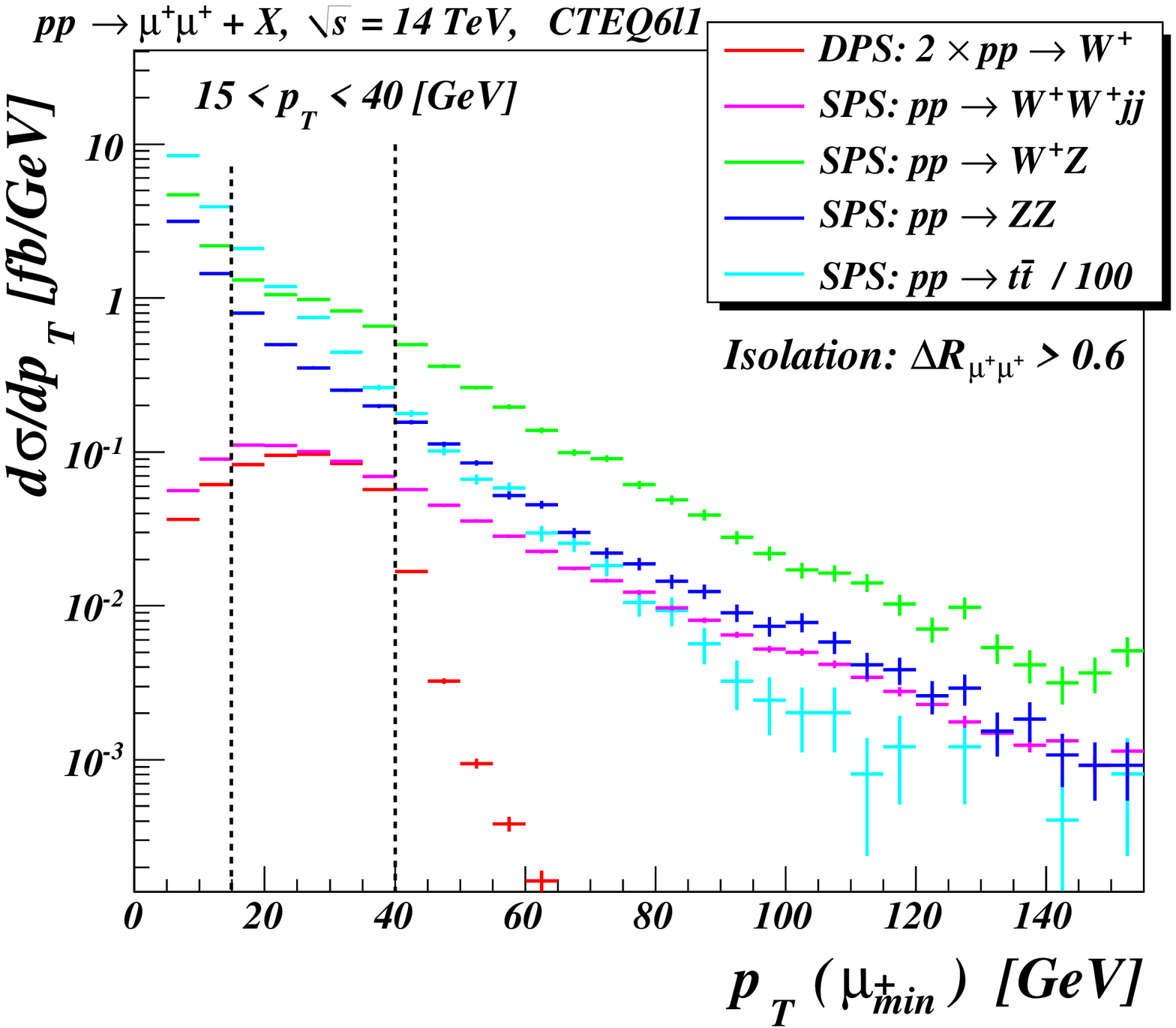}
   \end{center}
  \end{minipage}

  \begin{minipage}{.45\hsize}
   \begin{center}
     \includegraphics[width=6cm]{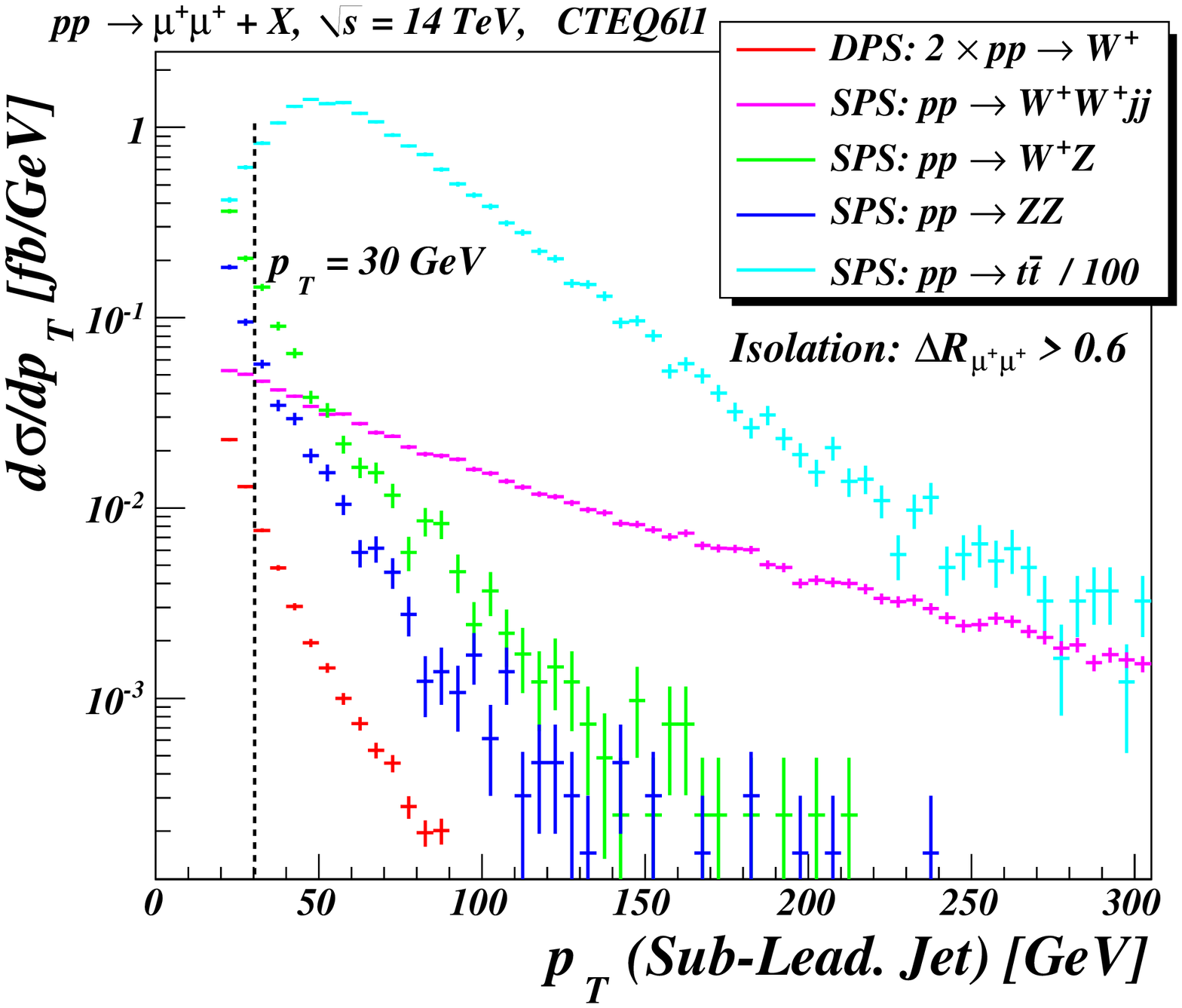}
   \end{center}
  \end{minipage}
 \end{tabular}
\caption{Differential cross section as a function of transverse
momentum of the second hardest positively charged muon (left plot) and of the sub-leading jet
(right plot). Distributions for $t\bar{t}$ process were reduced by 1/100.}
\label{fig:lep-jetPT}
\end{center}
\end{figure}

\begin{figure}[h!]
\begin{center}
 \begin{tabular}{cc}
  \begin{minipage}{.45\hsize}
   \begin{center}
     \includegraphics[width=6cm]{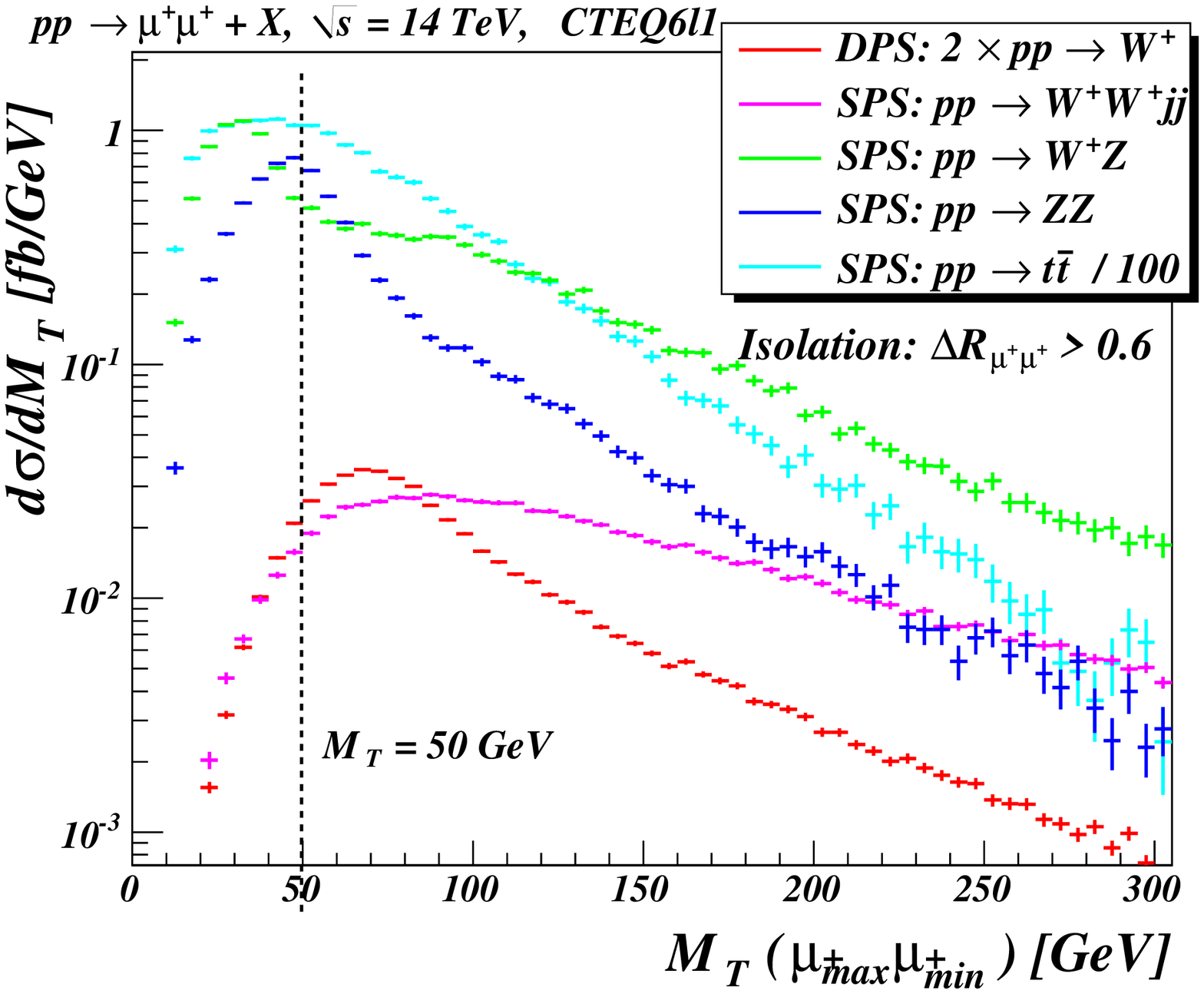}
   \end{center}
  \end{minipage}

  \begin{minipage}{.45\hsize}
   \begin{center}
     \includegraphics[width=6cm]{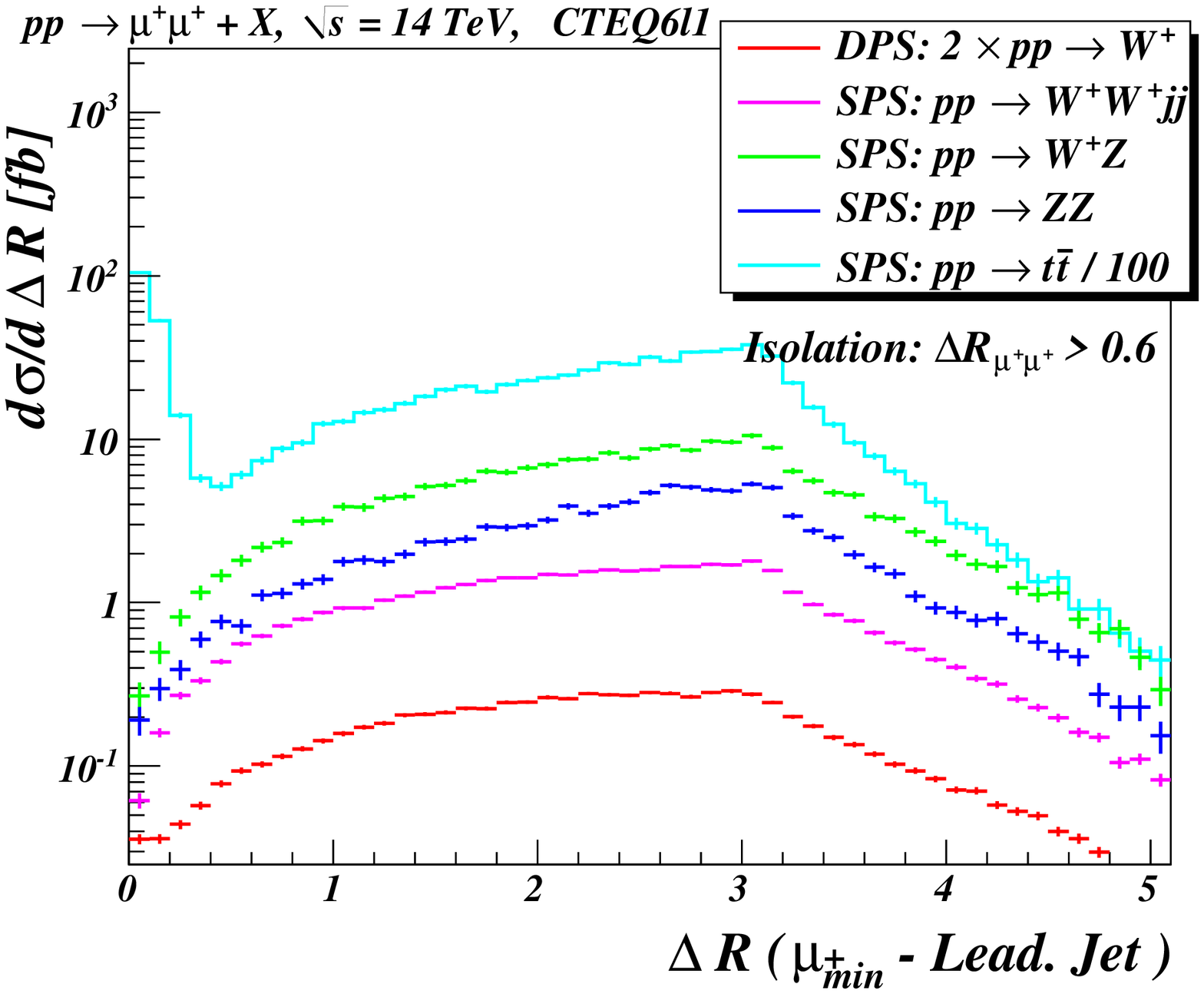}
   \end{center}
  \end{minipage}
 \end{tabular}
\caption{Differential cross section as a function of transverse mass of the positively
charged muon pair (left plot) and of the distance $\Delta R$ between the second hardest
positively charged muon and the leading jet (right plot).}
\label{fig:lep-jetR}
\end{center}
\end{figure}

The selection criteria are established on the basis of kinematics of two hardest positively
charged muons ($\mu^+_{max}$ and $\mu^+_{min}$) and of two hardest jets in the event.
Investigated variables are, for example, transverse momentum distributions of second hardest muon
and of the second hardest jet (see Fig. \ref{fig:lep-jetPT}). Also muon pair and muon-jet pair characteristics
(e.g. Fig. \ref{fig:lep-jetR}) are very important for the event analysis.

Especially heavy flavor quark production embodies a strong jet background which can be
used for the signal selection. Nevertheless, there is still a non-negligible signal
event fraction also with hard jets coming from QCD MPI or initial state radiations.
Simple isolation criteria between jets and muons can filter off a large amount
of $t\bar{t}$ processes.

The final selection criteria strongly depends on the transverse
momentum threshold of efficient muon detection and on the hadron
calorimeter acceptance. The lower $p_T$ limit of the muon detection
is considered, the more background can be filtered off. Similarly,
the larger calorimeter acceptance and the larger cone-like parameter
$R$ for the clustering algorithm is used, the more jets can be kept
in the event for further analysis. A very sensitive approach is
necessary because of the low signal cross section. The contemporary
stage of the study leads to the following $Final~selection$
criteria:

\begin{table}[h!]
\begin{center}
\begin{tabular}{c c c c c c c c c c}
$20~GeV$ & $<$ & $p_T(\mu^+_{max})$ & $<$ & $50~GeV$ & $15~GeV$ & $<$ & $p_T(\mu^+_{min})$ & $<$ & $40~GeV$ \\
\end{tabular}
\begin{tabular}{c c c c c c c c c c}
& $M (\mu^+_{max}\mu^+_{min})$ & $>$ & $20~GeV$ & &  & $M_T(\mu^+_{max}\mu^+_{min})$ & $>$ & $50~GeV$ & \\
& $p_T(\mu^-_{max})$ & $<$ & $5~GeV$ & & & $missing~E_T$ & $>$ & $20~GeV$ & \\
& $p_T(jet_{lead.})$ & $<$ & $40~GeV$ & & & $p_T(jet_{s-lead.})$ & $<$ & $30~GeV$ & \\
& $\Delta R_{\mu^+ jet}$ & $>$ & $0.4$ & & & $\Delta R_{\mu^+_{max}\mu^+_{min}}$ & $>$ & $0.6$ &
\end{tabular}
\end{center}
\end{table}

All the cross sections for individual contributions to the searched di-muon final
state are written in Table \ref{tab:sigma_summary}. $All~Events$ cathegory contains
all events where muons have to satisfy

\begin{equation}
p_T (\mu^+) > 5~GeV,~~|\eta (\mu^+)| < 2.5,~~\Delta R(\mu^+_{max}\mu^+_{min}) > 0.6.
\end{equation}

\noindent Double parton scattering here is completely overwhelmed by
background. Two main selections follow. First, the
$Negative~muon~veto$ rejected events containing a negatively charged
muon with $p_T (\mu^+) > 5~GeV$. Almost one third of the remaining
$W^+Z$ and $ZZ$ background can be further separated by going down to
$3~GeV$ $p_T$ cut without observable change of signal contribution.
Unfortunately, trigger efficiencies and other detector effects have
not been studied yet. Second substantial selection is the
$Jet~veto$. Here, events containing at least one jet with $p_T >
20~GeV$ are rejected. This selection filters significantly the
$W^+W^+jj$ and $t\bar{t}$ contributions off but keeps only $60\%$ of
the signal events. Therefore the $Final~selection$ releases this jet
cut a little bit and uses the isolation cut. Additionally, the
b-tagging is supposed to reduce the $t\bar{t}$ background at least
by $50\%$ but it is not involved here. Total cross section for the
searched di-muon final state is predicted to be around 3.66 $fb$
while the signal-to-background ratio is 0.35. Necessary integrated
luminosity recorded e.g. by the ATLAS detector has to be above 103
$fb^{-1}$ to reach the signal significance at least five standard
deviations above the background.

\begin{table}[h!]
\begin{center}
\begin{tabular}{l | c | c c c c}
$\sigma~[fb]$ & DPS Signal & \multicolumn{4}{|c}{SPS Background}\\
 & $W^+W^+$ & $W^+W^+jj$ & $W^+Z$ & $ZZ$ & $t\bar{t}$ \\
\hline
$All~events$           & 1.96 & 4.59 & 68.21 & 36.41 & 8.8$\times10^3$ \\
$Negative~muon~veto$   & 1.95 & 4.54 & 19.81 &  2.87 & 6.7$\times10^3$ \\

\footnotesize ~~~~survived    & \footnotesize 99\% & \footnotesize 99\% & \footnotesize 29\% & \footnotesize 8\%  & \footnotesize 76\%         \\

$Jet~veto$        & 1.18 & 0.14 & 46.48 & 25.74 & 13.66           \\

\footnotesize ~~~~survived    & \footnotesize 60\% & \footnotesize 3\% & \footnotesize 68\%  & \footnotesize 71\%  & \tiny$<$\footnotesize 1\%              \\

$Final~selection$  & 0.94 & 0.06 & 1.93 & 0.13 & 0.59              \\

\footnotesize ~~~~survived    & \footnotesize 48\% & \footnotesize 1\%  & \footnotesize 3\%  & \tiny $<$\footnotesize 1\%  & \tiny $<<$\footnotesize 1\%       \\
\hline
\end{tabular}
\caption{Summary of five studied processes characterized by production cross sections in
femto-barns as well as by the fractions of appropriate events
surviving the individual selections. }
\label{tab:sigma_summary}
\end{center}
\end{table}

\newpage

\section{Summary and Conclusions}

The importance of the double parton scattering was outlined in the
manner of the increasing energy and luminosity available at the LHC.
Relatively new Herwig++ generator incorporates the eikonal model of
hadron interactions and provides very useful tool for the estimation
of the particle behavior within the multiple parton scattering. This
study uses its model in order to generate proton-proton collisions
at $\sqrt{s} = 14~TeV$, where pair of positively charged gauge
bosons is produced via two independent quark-antiquark annihilations
occurring within the same collision. The muon decay channel is
studied in detail. The studied background processes include the single parton
scattering producing gauge boson pair with any charge combination or
heavy flavor quark pair. The latter is represented here by the
analysis of $t\bar{t}$ events. The background
processes were prepared using both MadGraph and Herwig++ programs in
order to study complete events including showering and
hadronization procedures.

Several mechanisms were studied in order to find the kinematical
region for the statistically reasonable measurement of the double
parton scattering. Roughly speaking, the strict event veto on the
negatively charged muons filters most of the single parton processes
producing at least one neutral gauge boson. Process $pp \rightarrow
W^+W^+jj$ and heavy flavor quark production are suppressed mostly by
selecting events without any hard jet (with $p_T > 40~GeV$). The
$Final~selection$ predicts the double parton scattering cross
section to be approximately 0.9 $fb$. Signal-to-background ratio
reaches 0.35. The true result will strongly depend on the detector
performance and on the trigger efficiencies that are not
incorporated in this study. The integrated luminosity is required to
be very large, of $\mathcal{O}(100fb^{-1})$. However, the LHC could
provide enough of statistics within a few years of full energy
operation.


\begin{thebibliography}{99}


\bibitem{PLB268_145}
J. Alitti et al. (UA2 Collaboration), Phys. Lett. B268 (1991) 145-154.

\bibitem{PRD56_3811}
F. Abe et al. (CDF Collaboration), Phys. Rev. D56 (1997) 3811.

\bibitem{EPJC58_639}
M. B\"ahr et al., Eur. Phys. J. C58 (2008) 639.

\bibitem{JHEP09_028}
J. Alwall et al., JHEP 0709 (2007) 028.

\bibitem{JHEP07_012}
J. Pumplin et al., JHEP 0207 (2002) 012.

\bibitem{PRD63_057901}
A. Del Fabbro and D.Treleani, Phys. Rev. D63 (2001) 057901.

\bibitem{PRD76_076006}
D. Treleani, Phys. Rev. D76 (2007) 076006.

\bibitem{EPJC71_1}
G. Aad et al. (ATLAS Collaboration), Eur. Phys.J. C71 (2011) 1.

\bibitem{JHEP04_063}
M. Cacciari and G. P. Salam, G. P., JHEP 0804 (2008) 063.

\bibitem{PLB641_57}
M. Cacciari, G. P. Salam and G. Soyez, Phys. Lett. B641 (2006) 57

\bibitem{CERN-OPEN-2008-020}
G. Aad et al. (ATLAS Collaboration), CERN-OPEN-2008-020, [arXiv:hep-ex/0901.0512]


\end{thebibliography}
\end{document}